\documentclass[review, 1p]{elsarticle}

\usepackage{amsmath,amssymb,bm}
\usepackage{lineno}

\journal{Physics Letters B}

\begin{document}

\begin{frontmatter}

\title{Two quasiparticle wobbling in the even-even nucleus $^{130}$Ba}

\author[PKU]{Y. K. Wang}
\author[NPU]{F. Q. Chen}
\author[PKU]{P. W. Zhao\corref{mycorrespondingauthor}}
\cortext[mycorrespondingauthor]{Corresponding author}
\ead{pwzhao@pku.edu.cn} 

\address[PKU]{State Key Laboratory of Nuclear Physics and Technology, School of Physics, Peking University, Beijing 100871, China}
\address[NPU]{School of Physical Science and Technology, Northwestern Polytechnical University, Xi'an 710129, China}

\begin{abstract}
Two newly observed bands built on a two-quasiparticle configuration in $^{130}$Ba have been investigated for the first time with the microscopic projected shell model. 
The experimental energy spectra and the available electromagnetic transition probabilities are well reproduced. 
The wobbling character of the higher band is revealed by the angular momentum projected wavefunctions via the \textit{K plot} and the \textit{azimuthal plot}. 
This provides the first strong microscopic evidence for wobbling motion based on a two-quasiparticle configuration in even-even nuclei. 
\end{abstract}

\begin{keyword}
Nuclear wobbling \sep Projected shell model \sep \textit{K plot} \sep \textit{Azimuthal plot}
\end{keyword}

\end{frontmatter}


\section{\label{sec1} Introduction}
The wobbling motion is one of the most intriguing quantum phenomena of a triaxial rotating nucleus, proposed by Bohr and Mottelson~\cite{Bohr1998}. It is a quantum analog to the motion of a free classical asymmetric top, whose rotation around the principal axis with the largest moment of inertia is usually energy favored and stable. 
The term ``stable'' here means that at slightly larger energies, the rotating axis would not be very far away from the space-fixed angular momentum vector, but instead, it executes harmonic precession oscillations about the space-fixed angular momentum. 
For quantal nuclear systems, these oscillations appear as equidistant excitations and, thus, the energy spectrum is a series of $\Delta I = 2\hbar$ rotational bands corresponding to increasing phonon quanta $n$ and alternating signature quantum numbers. 
Moreover, the $\Delta I = 1\hbar$ transitions between the bands with $n$ and $n+1$ phonons are collectively enhanced. 

A clear evidence for wobbling in this purely collective form, which is seen in all asymmetric top molecules, has not been found so far in nuclear systems. Instead, wobbling evidences have only been reported in odd-$A$ triaxial nuclei, e.g, $^{161}$Lu~\cite{Bringel2005Eur.Phys.J.A167}, $^{163}$Lu~\cite{Odegard2001Phys.Rev.Lett.5866,Jensen2002Phys.Rev.Lett.142503}, $^{165}$Lu~\cite{Schoenwaser2003Phys.Lett.B9}, $^{167}$Lu~\cite{Amro2003Phys.Lett.B197}, $^{167}$Ta~\cite{Hartley2009Phys.Rev.C41304}, $^{135}$Pr~\cite{Matta2015Phys.Rev.Lett.82501,Sensharma2019Phys.Lett.B170}, and $^{105}$Pd~\cite{Timar2019Phys.Rev.Lett.62501}, where either an odd proton or a neutron occupying a high angular momentum orbital is coupled to the triaxial rotor, and considerably influences the wobbling motion.  
As a result, the experimentally observed wobbling energies, i.e., energy differences between the wobbling bands, have been found to decrease with increasing spin, contrary to the behavior expected for even-even nuclei~\cite{Bohr1998}. 
This is interpreted as the so-called ``transverse wobbling''~\cite{Frauendorf2014Phys.Rev.C14322}, where the odd-nucleon angular momentum alignment is assumed to be frozen and perpendicular to the axis with the maximal moment of inertia. 
This interpretation stimulates great theoretical interests to clarify the modified wobbling mode in odd-$A$ nuclei using different models~\cite{Chen2014Phys.Rev.C44306,Tanabe2017Phys.Rev.C64315,Raduta2017Phys.Rev.C54320,Shimada2018Phys.Rev.C24318,Budaca2018Phys.Rev.C24302}. 

Only a few indications for even-even wobbler have been reported. For instance, in $^{112}$Ru~\cite{Hamilton2010Nucl.Phys.A28}, the ground band, and the odd and even spin members of the ``$\gamma$-band'' are proposed as the zero-, one-, and two-phonon wobbling bands, respectively. 
However, this wobbling interpretation is not very solid because no electromagnetic transition data were reported. 
A very interesting example of wobbling in even-even nuclei is the recently reported band structure in $^{130}$Ba~\cite{Petrache2019Phys.Lett.B241}, where a pair of bands with even and odd spins, labeled S1 and S1', were interpreted as the zero- and one-phonon wobbling bands~\cite{Chen2019Phys.Rev.C61301}. 
It should be mentioned that the configuration of these two bands are built on two aligned protons in the bottom of the $h_{11/2}$ shell. 
Therefore, the wobbling excitation in $^{130}$Ba is not in a purely collective form, but in the presence of two aligned particles. 

On the theoretical side, nuclear wobbling bands have been extensively studied with the triaxial particle-rotor model~\cite{Frauendorf2014Phys.Rev.C14322,Matta2015Phys.Rev.Lett.82501,Timar2019Phys.Rev.Lett.62501}. However, such analyses are all phenomenological and are fitted to the data in one way or another. 
There are also many efforts to extend the microscopic cranking mean-field model to study the wobbling motion.
The cranking mean-field model yields only the lowest state for a given configuration and, thus, one has to go beyond the mean-filed level to describe the wobbling excitations. 
This has been done by incorporating the quantum correlations by means of random phase approximation (RPA)~\cite{Matsuzaki2002Phys.Rev.C41303,Nakatsukasa2016Phys.Scr.73008} or by the angular momentum projection methods~\cite{Oi2000Phys.Lett.B53,Shimada2018Phys.Rev.C24318}.

The projected shell model (PSM) carries out the shell-model configuration mixing based on Nilsson mean field with the angular momentum projection technique~\cite{Hara1995Int.J.Mod.Phys.E637}. The implementation can also be rooted on the self-consistent relativistic~\cite{Zhao2016Phys.Rev.C41301} and nonrelativistic~\cite{Konieczka2018Phys.Rev.C34310} density functional theories.
The PSM was used to understand the wobbling motion in $^{135}$Pr~\cite{Sensharma2019Phys.Lett.B170}, where the observed energy spectra and electromagnetic transitions for the wobbling bands are well reproduced. 
However, an illustration for the underlying wobbling geometry of the angular momentum was missing. 
The difficulty lies in the fact that the angular momentum geometry is defined in the intrinsic frame, while the angular momentum projected wavefunctions are written in the laboratory frame. 
In Ref.~\cite{Chen2017Phys.Rev.C51303}, focusing on the chiral doublet bands in triaxial nuclei~\cite{Frauendorf1997Nucl.Phys.A131}, the \textit{K plot} and the \textit{azimuthal plot} are introduced to illustrate the chiral geometry with the angular momentum projected wavefunctions. 

In this work, we report a microscopic investigation on the recently observed two-quasiparticle bands S1 and S1' of $^{130}$Ba with the PSM. 
This is the first example of wobbling motion based on a two-quasiparticle configuration. 
In particular, the influence of the two quasiparticles on the angular momentum geometry of the wobbling bands is illustrated in terms of the \textit{K plot} and the \textit{azimuthal plot}.

\section{\label{sec2} Theoretical Framework}
The framework of the PSM starts from the standard pairing plus quadrupole Hamiltonian~\cite{Ring1980},
\begin{equation}\label{eq1}
  \hat{H} = \hat{H}_0 - \frac{\chi}{2}\sum_{\mu}\hat{Q}^\dag_\mu\hat{Q}_\mu - G_M\hat{P}^\dag\hat{P} - G_Q\sum_{\mu}\hat{P}^\dag_\mu\hat{P}_\mu,
\end{equation}
which includes a spherical single-particle shell model Hamiltonian, a quadrupole-quadrupole interaction, a monopole pairing interaction, and a quadrupole pairing interaction.
The intrinsic vacuum state $|\Phi_0\rangle$ can be calculated by the following variational equation,
\begin{equation}\label{eq2}
  \delta\langle\Phi_0|\hat{H}-\lambda_n\hat{N}-\lambda_p\hat{Z}|\Phi_0\rangle = 0,
\end{equation}
with the Lagrange multipliers $\lambda_n$ and $\lambda_p$ determined by the neutron number $N$ and proton number $Z$, respectively.

Based on the intrinsic vacuum state $|\Phi_0\rangle$, the two quasiparticle states $|\Phi_\kappa\rangle$ for even-even nuclei can be constructed with
\begin{equation}\label{eq3}
  |\Phi_\kappa\rangle \in \{\hat{\beta}^\dag_{\nu_i}\hat{\beta}^\dag_{\nu_j}|\Phi_0\rangle, \hat{\beta}^\dag_{\pi_i}\hat{\beta}^\dag_{\pi_j}|\Phi_0\rangle\},
\end{equation}
where $\hat{\beta}^\dag_\nu$ and $\hat{\beta}^\dag_\pi$ are the quasiparticle creation operators for neutron and proton, respectively.
The rotational symmetry of the intrinsic states $|\Phi_\kappa\rangle$ can be restored by the projection $\{\hat{P}^I_{MK}|\Phi_\kappa\rangle\}$,
in which $\hat{P}^I_{MK}$ denotes the three-dimensional angular momentum projection operator~\cite{Ring1980}. 

The Hamiltonian~(\ref{eq1}) is diagonalized in the space consisting of the projected two quasiparticle states and the vacuum,  and this leads to the Hill-Wheeler equation,
\begin{equation}\label{eq6}
  \sum_{\kappa'K'}\{\langle\Phi_{\kappa}|\hat{H}\hat{P}^I_{KK'}|\Phi_{\kappa'}\rangle-E^{I\sigma}\langle\Phi_{\kappa}|\hat{P}^I_{KK'}|
  \Phi_{\kappa'}\rangle\}f^{I\sigma}_{K'\kappa'} = 0,
\end{equation}
where $\sigma$ labels different eigenstates with the same spin $I$.
The norm matrix element $\mathcal{N}_I(K,\kappa;K',\kappa') = \langle\Phi_\kappa|\hat{P}^I_{KK'}|\Phi_{\kappa'}\rangle$ and the energy kernel $\mathcal{H}_I(K,\kappa;K',\kappa') = \langle\Phi_\kappa|\hat{H}\hat{P}^I_{KK'}|\Phi_{\kappa'}\rangle$ can be calculated by using the Pfaffian algorithm~\cite{Bertsch2012Phys.Rev.Lett.42505,Hu2014Phys.Lett.B162}. 

By solving the Hill-Wheeler equation (\ref{eq6}), one can obtain the eigenvalues $E^{I\sigma}$ and the corresponding eigenfunctions
\begin{equation}\label{eq7}
  |\Psi^\sigma_{IM}\rangle = \sum_{K\kappa} f^{I\sigma}_{K\kappa}\hat{P}^I_{MK}|\Phi_\kappa\rangle,
\end{equation}
with which the electromagnetic transitions can be calculated.
It is known that the projected basis $\{\hat{P}^I_{MK}|\Phi_\kappa\rangle\}$ are not orthogonal and therefore, the coefficients $f^{I\sigma}_{K\kappa}$ in Eq.~(\ref{eq7}) should not be understood as probability amplitudes. 
However, one can construct the orthogonal and normalized collective wavefunctions~\cite{Ring1980}
\begin{equation}\label{eq8}
  g^{I\sigma}(K,\kappa) = \sum_{K'\kappa'}\mathcal{N}_I^{1/2}(K,\kappa; K',\kappa')f^{I\sigma}_{K'\kappa'},
\end{equation}
which are interpreted as probability amplitudes and are used to construct the \textit{K plot} and the \textit{azimuthal plot}~\cite{Chen2017Phys.Rev.C51303}.
The \textit{K plot} is defined as the probability distributions of the components of the angular momentum on the three axes of the intrinsic frame,
\begin{equation}\label{eq9}
  p^{I\sigma}(|K|) = \sum_\kappa \left|g^{I\sigma}(K,\kappa)\right|^2 + \left|g^{I\sigma}(-K,\kappa)\right|^2.
\end{equation}
The \textit{azimuthal plot} is defined as the probability distributions of the polar and the azimuthal angles $(\theta,\phi)$ of the angular momentum in the intrinsic frame,
\begin{equation}\label{eq10}
  \mathcal{P}(\theta,\phi) = \sum_\kappa\int_0^{2\pi} d\psi'\left|G^{II}(\psi',\theta,\pi-\phi,\kappa)\right|^2,
\end{equation}
where $\theta$ denotes the angle between the total angular momentum and the long (\textit{l}) axis, and $\phi$ denotes the angle between the projection of the total angular momentum on the intermediate-short (\textit{i-s}) plane and the \textit{i} axis.
The integrand $G^{II}(\psi',\theta,\pi-\phi)$ reads,
\begin{equation}\label{eq11}
  G^{II}(\psi',\theta,\pi-\phi,\kappa) = \sqrt{\frac{2I+1}{8\pi^2}}\sum_K g^I(K,\kappa)D^{I\ast}_{IK}(\psi',\theta,\pi-\phi).
\end{equation}

In the present calculations for the observed bands S1 and S1' in Ref.~\cite{Petrache2019Phys.Lett.B241}, both the proton and neutron orbitals are taken from the $N = 3,\, 4,\, 5$ major shells.
The monopole pairing strength, $G_M = 0.15$ MeV for neutron and $G_M = 0.16$ MeV for proton, is determined by the odd-even mass differences. 
The strength of quadrupole pairing interaction $G_Q$ is assumed to be $0.2\, G_M$; similar to many other calculations~\cite{Sun1994Phys.Rev.Lett.3483,Sun1996Phys.Rep.375}. 
The strength of quadrupole force $\chi$ is connected with the quadrupole deformation $(\beta,\gamma)$ by the self-consistent relation as shown in Ref.~\cite{Hara1995Int.J.Mod.Phys.E637}.
The deformation parameters $(\beta,\gamma)$ are chosen as $(0.24,30^\circ)$, which is similar to the results given by the calculations of the cranking covariant density functional theory~\cite{Zhao2011Phys.Rev.Lett.122501,Zhao2011Phys.Lett.B181,Zhao2015Phys.Rev.Lett.22501} with PC-PK1~\cite{Zhao2010Phys.Rev.C54319}.
Since the configuration of bands S1 and S1' is assigned as $\pi(h_{11/2})^2$~\cite{Petrache2019Phys.Lett.B241}, only the proton orbitals in the $h_{11/2}$ shell are used to construct the configuration space. 

\section{\label{sec3} Results and discussion}

\begin{figure}[!htbp]
  \centering
  \includegraphics[width=0.4\textwidth]{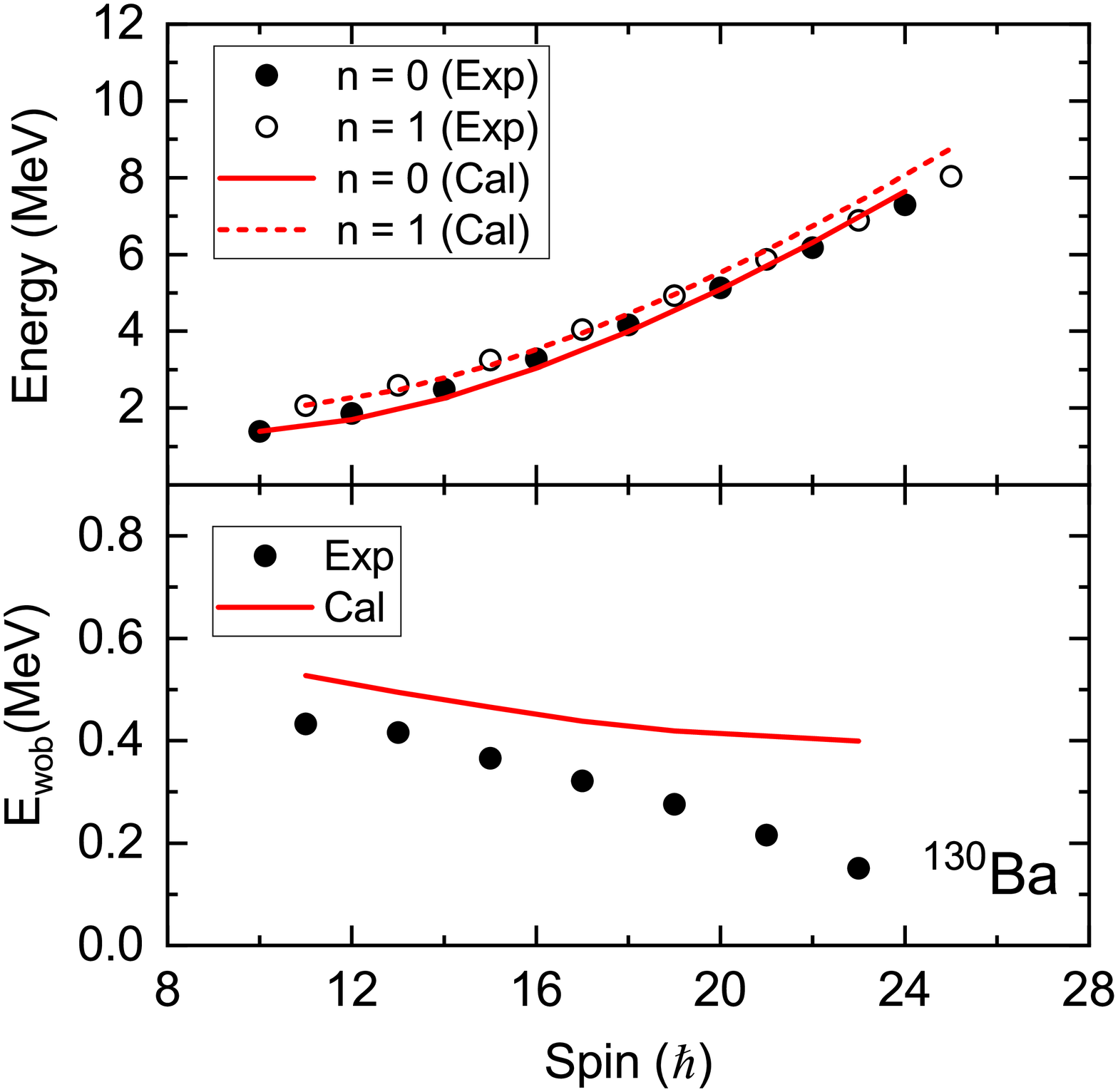}
  \caption{(Color online) Calculated energy spectra (top) for the zero- ($n = 0$) and one-phonon ($n = 1$) bands as well as the wobbling energy (bottom), in comparison with data~\cite{Petrache2019Phys.Lett.B241}.}
  \label{Energy}
\end{figure}

In the upper panel of Fig.~\ref{Energy}, the calculated excitation energies with respect to the bandhead ($I=10\hbar$) for the zero- ($n=0$) and one-phonon ($n=1$) states are presented in comparison with data~\cite{Petrache2019Phys.Lett.B241}. 
It is seen that the observed excitation energies are well reproduced, in particular for the lower spin states. For the higher spin states, the calculated energies are slightly higher than data, but this could be improved by including more configurations in the model space. 
The wobbling energies $E_{\mathrm{wob}}$, defined as 
\begin{equation}
E_{\mathrm{wob}} (I) = E_{n=1}(I) - \left[ E_{n=0}(I+1)+E_{n=0}(I-1) \right]/2,
\end{equation}
were calculated from the energy spectra and are shown in the lower panel of Fig.~\ref{Energy}. 
The calculated results are in good agreement with data, and in particular, the decreasing tendency of the wobbling energy as a function of angular momentum is presented. 
This has been suggested as the hallmark of the transverse wobbling in Ref.~\cite{Frauendorf2014Phys.Rev.C14322}. 

\begin{figure}[htbp]
  \centering
  \includegraphics[width=0.4\textwidth]{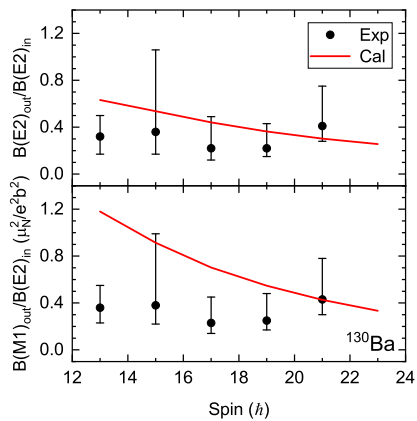}
  \caption{(Color online) Calculated transition probability ratios $B(E2)_{\rm out}/B(E2)_{\rm in}$ (top) and $B(M1)_{\rm out}/B(E2)_{\rm in}$ (bottom) for the transitions from the one-phonon ($n = 1$) band to the zero-phonon ($n = 0$) band in comparison with data available~\cite{Petrache2019Phys.Lett.B241}.}
  \label{Transition}
\end{figure}

The experimental and theoretical transition probability ratios $B(M1)_{\rm out}/B(E2)_{\rm in}$ and $B(E2)_{\rm out}/B(E2)_{\rm in}$ for the transitions from the one-phonon band to the zero-phonon band are shown in Fig.~\ref{Transition}.  
The calculated results agree well with the data. 
Different from an ideal wobbler, here the values of $B(M1)_{\rm out}/B(E2)_{\rm in}$ are comparable with the $B(E2)_{\rm out}/B(E2)_{\rm in}$ ones. 
This is attributed to the fact that two quasiparticles in the $h_{11/2}$ shell are involved in the configurations, which enlarges the $M1$ matrix elements.
Therefore, the $E2$ component of the $\Delta I=1$ transitions is not expected to be that dominate like in other wobbling bands~\cite{Matta2015Phys.Rev.Lett.82501}.   

\begin{figure*}[htbp]
  \centering
  \includegraphics[width=0.7\textwidth]{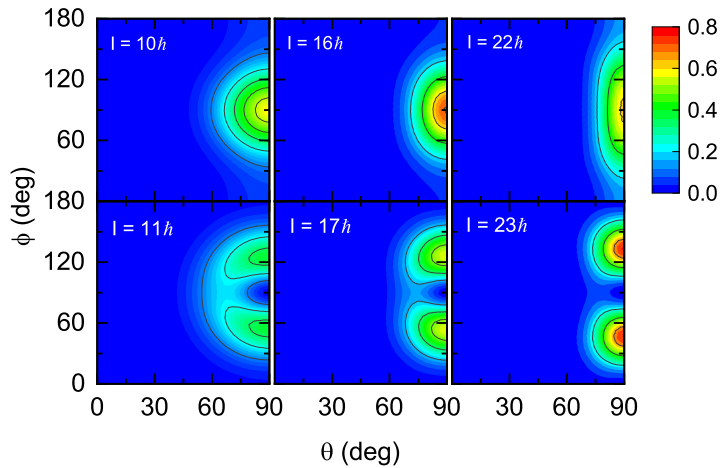}
  \caption{(Color online) The \textit {azimuthal plots}, i.e., probability distribution profiles for the orientation of the angular momentum on the intrinsic $(\theta,\phi)$ plane for the zero- (even spin) and the one- (odd spin) phonon band at several selected angular momenta.}
  \label{Aplot}
\end{figure*}

In order to examine the orientation of the angular momentum on the intrinsic $(\theta, \phi)$ plane, Fig.~\ref{Aplot} depicts the \textit{azimuthal plots}, i.e., the profiles $\mathcal{P}(\theta, \phi)$ for the zero- (even spin) and one- (odd spin) phonon bands at several selected angular momenta. 
In the $\theta$ direction, for all states, the distributions $\mathcal{P}(\theta, \phi)$ are mainly concentrated around $\theta \sim 90^\circ$, and this means that the angular momentum locates mainly in the plane of short and intermediate axes. 
For both the zero- and one-phonon states, the distributions in the $\theta$ direction are more diffuse at lower spin states.
In the $\phi$ direction, however, distinct patterns are found for the zero- and one-phonon states.  
For the zero-phonon states with $I = 10,\,16,\,22 \hbar$, the distributions $\mathcal{P}(\theta, \phi)$ are mainly concentrated around $\phi\sim90^\circ$, and they become more and more diffuse with the increasing spin. 
For the one-phonon states with $I = 11,\,17,\,23 \hbar$, the angular momenta orientate equally at two directions with $\phi\sim 120^\circ$ and $\phi\sim 60^\circ$, respectively, and the separation of the two directions become larger with the increasing spin.
In contrast to the maximum at $\phi \sim 90^\circ$ for the distribution $\mathcal{P}(\theta, \phi)$ of the zero-phonon states, there is a minimum at $\phi \sim 90^\circ$ in the distribution for the one-phonon states.
Therefore, the distribution in the $\phi$ direction is symmetric for the zero-phonon states and is antisymmetric for the one-phonon states.
This pattern is consistent with the expectation of a wobbling motion, i.e., the precession of the total angular momentum around the short axis.  

\begin{figure*}[htbp]
  \centering
  \includegraphics[width=0.7\textwidth]{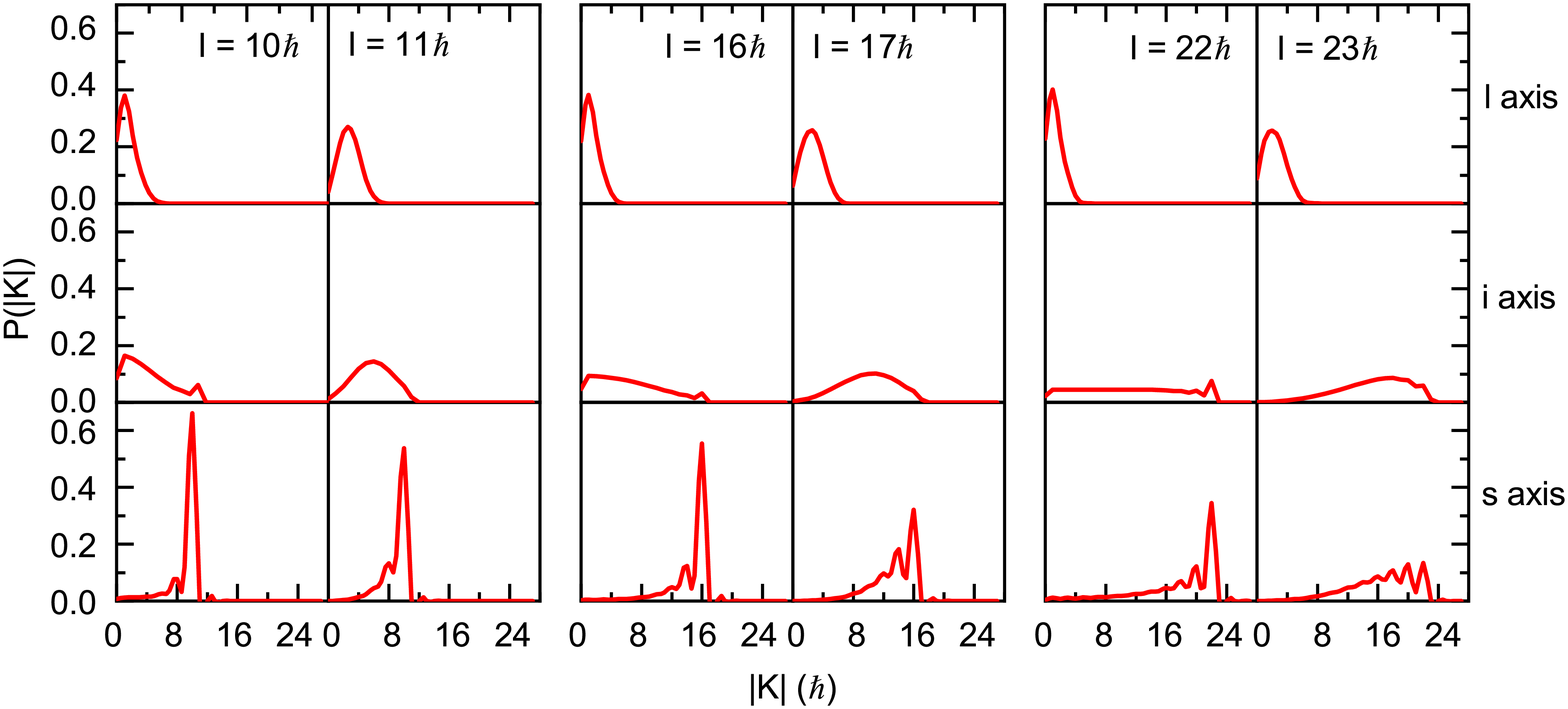}
  \caption{(Color online) The \textit{K plot}, i.e., the $K$ distributions of angular momenta on the three principle axes for the zero- (even spin) and one- (odd spin) phonon band at several selected angular momenta.
  The $K$ distributions on the \textit{long}, \textit{intermediate}, and \textit{short} axes are shown in the first, second, and third rows, respectively.}
  \label{Kplot}
\end{figure*}

In Fig.~\ref{Kplot}, the \textit{K plot}, i.e., the $K$ distributions of angular momenta on the three principle axes for the zero- (even spin) and one- (odd spin) phonon bands are depicted at several selected angular momenta.
At $I = 10\hbar$, the angular momentum is mainly along the short axis, due to the alignment of the two quasiparticles in the $h_{11/2}$ shell. Therefore, it is seen in Fig.~\ref{Kplot} that the probability is peaked at very small $K$ values on the \textit{long} and \textit{intermediate} axes, while at $K\sim 10\hbar$ on the \textit{short} axis. 
At $I = 11\hbar$, the $K$ distributions changes mainly on the \textit{intermediate} axis, i.e., the remarkable probability at $K_i\sim 0$ for $I = 10 \hbar$ vanishes.   
This indicates that a symmetric wavefunction with respect to $K_i \sim 0$ at $I = 10 \hbar$ changes to an antisymmetric one at $I = 11\hbar$. 
It is consistent with the pattern shown in the \textit {azimuthal plots} (see Fig.~\ref{Aplot}), and reflects the fact that the states at $I = 10\hbar$ and $11\hbar$ correspond to the zero- and one-phonon states, respectively.  

Increasing the spin to $I = 16\hbar$ and $22\hbar$, the $K$ distributions on the \textit{long} axis barely change, which means that the angular momentum grows very little in the  \textit{long}-axis direction. 
Comparing the $K$ distributions on the other two axes, it is found that the angular momentum is generated mainly in the short axis. 
Though the two $h_{11/2}$ quasiparticles contribute only about $10\hbar$ in the short axis, the peaks of $K$ distributions on the short axis can reach $16\hbar$ and $22\hbar$ for the $I = 16\hbar$ and $22\hbar$ states, respectively. 
This indicates that the scenario of the transverse wobbling~\cite{Frauendorf2014Phys.Rev.C14322} is realized in the present microscopic calculations.  
The $K$ distributions on the \textit{intermediate} axis become very broad for the $I = 16\hbar$ and $22\hbar$ states, while the significant probabilities at $K_i\sim 0$ remain.
This feature, together with the vanishing probabilities at $K_i\sim 0$ for the $I = 17 \hbar$ and $23\hbar$ states, reflects again the wobbling character of the odd-spin states.  

For the $I = 23 \hbar$ state, it can be seen that the components of the angular momentum on the \textit{intermediate} and \textit{short} axes are comparable. This might indicate that $I = 23 \hbar$ is very close to the so-called critical spin, and above this spin, the transverse wobbling would not be stable and the tilted axis rotation may appear. 

\section{\label{sec4} Summary}
In summary, a pair of newly observed bands built on a two-quasiparticle configuration in $^{130}$Ba have been investigated with the microscopic projected shell model. 
The experimental energy spectra, energy differences between the two bands, as well as the available electromagnetic transition probabilities are well reproduced. 
Different from an ideal wobbler, the values of $B(M1)_{\rm out}/B(E2)_{\rm in}$ are comparable with the $B(E2)_{\rm out}/B(E2)_{\rm in}$ ones due to the involvement of the two quasiparticles in the configurations.
Nevertheless, the wobbling character of the higher band can be demonstrated in terms of the \textit{K plot} and the \textit{azimuthal plot}. This provides the first strong microscopic evidence of wobbling motion based on a two-quasiparticle configuration in even-even nuclei. 

\section*{Acknowledgments}
This work was partly supported by the National Key R\&D Program of China (Contract No. 2018YFA0404400 and No. 2017YFE0116700), 
the National Natural Science Foundation of China (Grants No. 11621131001, No. 11875075, No. 11935003, and No. 11975031), and the Laboratory Computing Resource Center at Argonne National Laboratory.



\end{document}